\documentclass[aps,pra,reprint,nofootinbib,twocolumn,superscriptaddress,showpacs,showkeys,longbibliography]{revtex4-1}
\usepackage{eurosym}
\usepackage{amsmath,amssymb,amstext}
\usepackage{ulem} 
\usepackage[usenames,dvipsnames]{color}
\usepackage{graphicx}
\usepackage{braket}
\usepackage{natbib}
\usepackage{comment}
\usepackage{dcolumn}
\usepackage{color}
\usepackage{wasysym}

\usepackage[colorlinks,bookmarks=false,citecolor=blue,linkcolor=red,urlcolor=blue]{hyperref}

\begin{document}
\title{Quench Dynamics and Stability of Dark Solitons in Exciton--Polariton
Condensates}

\author{Chunyu jia}
\affiliation{College of Physical Science and Technology, Bohai University, Jinzhou 121013, China}

\author{Zhaoxin Liang}
\email{The corresponding author: zhxliang@zjnu.edu.cn}
\affiliation{College of Physics and Electronic Information Engineering, Zhejiang Normal University, Jinhua, 321004, China}

\date{\today}

\begin{abstract}
Exciton--polariton condensates (EPCs) have emerged as a paradigmatic platform for investigating nonequilibrium quantum many-body phenomena, particularly due to their intrinsic open-dissipative nature and strong nonlinear interactions governed by the interplay between stimulated scattering and reservoir-mediated damping. 
Recent advances in Feshbach resonance engineering now enable precise tuning of interaction strengths, opening new avenues to explore exotic nonlinear excitations in these driven-dissipative systems.
In this work, we systematically investigate the quench dynamics and stability of dark solitons in repulsive one-dimensional EPCs under sudden parameter variations in both nonlinear interaction strength $g$ and pump intensity $P$. 
Through a Hamiltonian variational approach that incorporates reservoir damping effects, we derive reduced equations of motion for soliton velocity evolution that exhibit remarkable qualitative agreement with direct numerical simulations of the underlying open-dissipative Gross--Pitaevskii equation. 
Our results reveal three distinct dynamical regimes: (i) stable soliton propagation at intermediate pump powers, (ii) velocity-dependent soliton breakup above critical pumping thresholds, and (iii) parametric excitation of soliton trains under simultaneous interaction quenches. 
These findings establish a quantitative framework for understanding soliton dynamics in nonresonantly pumped EPCs, with implications for quantum fluid dynamics and nonequilibrium Bose--Einstein condensates.
\end{abstract}

\maketitle

\section{Introduction}
Exciton--polariton condensates (EPCs) in semiconductor microcavities represent a highly promising platform for investigating fundamental quantum many-body problems and nonequilibrium phenomena. Owing to their inherent nonequilibrium character, polariton condensates have garnered significant attention from the physics community over the past two decades. As~hybrid quasiparticles, exciton--polaritons exhibit properties of composite bosons, arising from the strong coupling between quantum-well excitons and cavity photons. The~photonic component endows polaritons with a low effective mass and a readily excitable nature, while the excitonic component mediates interactions between polaritons~\cite{Baumberg_Spontaneous_2008, Yiling_Dynamics_2022}. Benefiting from these synergistic attributes inherited from their constituent components, polaritons can more readily form macroscopic coherent quantum states—namely, exciton--polariton condensates—without requiring the ultralow temperatures necessary for conventional atomic condensates. To~date, multiple experimental groups have achieved exciton--polariton condensation at elevated temperatures~\cite{kasprzak_boseeinstein_2006, balili_bose-einstein_2007, deng_spatial_2007, amo_collective_2009, schneider_electrically_2013} and~even at room temperature~\cite{schneider_electrically_2013, christopoulos_room-temperature_2007}. Collectively, the~experimental realization of exciton--polariton condensates has inaugurated a new frontier of research, where the study of nonlinear excitations and nonequilibrium dynamics remains a persistent and captivating~focus.

Soliton formation in atomic Bose--Einstein condensates (BECs) is classically understood to arise from the balance between dispersion and nonlinear interactions~\cite{ruprecht_time-dependent_1995,burger_dark_1999}. Unlike conservative systems (e.g., atomic BECs), where solitons can persist indefinitely, exciton--polariton solitons exhibit finite lifetimes due to the rapid radiative decay inherent to their hybrid quasiparticle nature. Consequently, sustaining a stationary soliton state requires continuous pumping to compensate for polariton losses. These nonlinear excitations display heightened instability in higher-dimensional systems—specifically, in~two or three dimensions, solitons readily decay into vortex states. Such dynamics have made them a focal point of extensive research. Notable experimental and theoretical studies include reports of solitons~\cite{WLZhang_Coupled_2012,xue_creation_2014,Sich_Soliton_2016,Kartashov_latticesolitons_2016,Walker_Dark_2017,xu_darkbright_2019,Septembre_Soliton_2024,HuJunwei_Darksolitoncloning_2024} and vortices~\cite{krizhanovskii_effect_2010,lagoudakis_quantized_2008,ma_creation_2017,rubo_half_2007} in microcavity polariton condensates. In~particular, Refs.~\cite{cuevas_nonlinear_2011,pinsker_approximate_2015} have specifically examined dark solitons in polariton condensates, analyzing their behavior under resonant and nonresonant pumping configurations,~respectively.

Distinct nonlinear excitations such as bright and dark solitons emerge from competing interaction regimes in quantum many-body systems. While attractive interactions (characterized by negative s-wave scattering lengths $a<0$) stabilize bright solitons~\cite{wang_dark_2003}, repulsive interactions promote dark soliton formation. Feshbach resonances provide an unparalleled platform for investigating these nonlinear phenomena through precise manipulation of interaction strengths via magnetic field tuning of scattering lengths~\cite{timmermans_feshbach_1999,inouye_observation_1998}. Recent advances demonstrate that quenching interaction strengths using Feshbach resonances can induce dynamical phase transitions, as~evidenced by dark soliton splitting in atomic Bose--Einstein condensates when crossing critical interaction thresholds~\cite{gamayun_fate_2015}. This approach has further revealed rich nonlinear dynamics in systems ranging from ultracold atoms~\cite{chen_observation_2020} to exciton--polariton condensates~\cite{drescher_real-space_2019}, where parametric amplification and soliton trains emerge under controlled interaction quenches~\cite{deng_tuning_2017}. The~interplay between Feshbach-tuned interactions and nonequilibrium dynamics continues to unlock new regimes in quantum fluid~behavior.

Soliton splitting represents a fundamental nonlinear phenomenon that manifests across diverse physical systems, including optical fibers~\cite{Hamid_Solitonsplitting_1997,porsezian_soliton_2015} and~BECs~\cite{wales_splitting_2020,Li_Merging_2021}.
In photonic systems, higher-order dispersion or nonlinear effects can trigger soliton splitting.
This phenomenon is typically accompanied by the formation of quantum shock waves, characterized by spectral broadening and wave-breaking effects. 
Quantum shock waves have emerged as a distinct nonlinear phenomenon attracting significant research attention. 
One formation mechanism involves quantum coherence between the background and wavepackets~\cite{Simmons_What_2020}. 
These shock waves exhibit characteristic abrupt changes in physical properties across the wavefront~\cite{jia_quenched_2021}. 
Recent theoretical research demonstrates that controlled generation of shock waves in polariton condensates can be achieved through modulation of external potentials and incoherent pumping~\cite{wang_generation_2024}.
 Shock waves have not only been observed in BECs~\cite{Kulikov_Shock_2003,Damski_Formation_2004,Kamchatnov_Dissipationless_2004,Hoefer_Dispersive_2006} but~were found in optics~\cite{wan_dispersive_2007,Ghofraniha_Shock_2012,nuno_vectorial_2019},  water waves~\cite{Jang_Shock_2023,Yin_Investigation_2023}, and~even Rydberg atomic gases~\cite{Hang_Accessing_2023,Qin_Shock-wave_2024}.
Despite these amazing results, the~physics of quantum dynamics and nonlinear excitations using resonance techniques is still not completely understood. 
So far, there have not been studies on the quench dynamics and stability of dark solitons in polariton~condensates.

In this work, we are motivated to launch a systematic investigation of the quench dynamics and stability of dark solitons in repulsive, one-dimensional EPCs subjected to the sudden quenches of both the nonlinear interaction parameter and the laser pump strength. {Specifically, when the interaction strength undergoes an abrupt change, the~initial soliton splits and generates a pair of faster asymmetric solitons}. To~this end, by~adopting a dissipative Gross--Pitaevskii description of EPCs, we investigate the static and dynamical properties of dark soliton. First,  the~equation of motion for the center of mass of the dark soliton's center is derived  analytically by using the Hamiltonian approach. The~resulting equation captures how the combination of the quench and the open-dissipative character can affect the properties of the dark soliton. Further numerical solutions are designed to study the dynamics of the dark soliton by sudden quenches of both the nonlinear interaction parameter and the laser pump strength, which are in good agreement with the analytical~results.

The present paper is structured as follows. In Section 2, we introduce the theoretical framework and derive the effective Gross--Pitaevskii equation under the fast reservoir limit. Next, in Section 3, we investigate the quench dynamics of dark solitons within exciton--polariton condensates, presenting results from both analytical treatments and direct numerical simulations. Finally, in Section 4, we summarize our key~findings.


\section{The Theoretical~Model } \label{Sec2}

In the present work, we investigate the dynamics of dark solitons in polariton condensates using diverse quench engineering.
To this end, we consider an incoherently far off-resonantly pumped exciton–polariton condensate in a one-dimensional (1D) setting. 
In the mean-field approximation, the~behavior of the polariton condensate is governed by the generalized open-dissipative Gross--Pitaevskii (GP) equation, 
with the condensate wave function $\psi\left(x, t\right)$ coupled to the hot exciton reservoir density $n_{R}\left(x, t\right)$~\cite{wouters_excitations_2007,carretero-gonzalez_kortewegvries_2017,Carusotto2013}
so that
\begin{eqnarray}
i\hbar\frac{\partial\psi}{\partial t} & = & -\frac{\hbar^{2}}{2m}\frac{\partial^2}{\partial x^{2}}\psi+g_{C}\left|\psi\right|^{2}\psi \nonumber\\
&+&\left[g_{R}n_{R}+\frac{i\hbar}{2}\left(Rn_{R}-\gamma_{C}\right)\right]\psi, \label{eq.1} \\ 
\frac{\partial n_{R}}{\partial t} & = & P-\left(\gamma_{R}+R\left|\psi\right|^{2}\right)n_{R}. \label{eq.2}
\end{eqnarray}
Before  proceeding, we remark that the detailed derivation of Equations~(\ref{eq.1}) and (\ref{eq.2}) can be found in the review work of Ref.~\cite{Carusotto2013}.

In Equations~(\ref{eq.1}) and (\ref{eq.2}), $m$ is the effect mass of the polariton, 
$g_{C}$ is proportional to the polariton--polariton interaction strength,
$g_{R}$ is proportional to the interaction strength between the hot exciton reservoir and polariton condensate, 
the condensate polaritons are continuously replenished from the hot exciton reservoir at a rate $R$ , 
$\gamma_{C}$ and $\gamma_{R}$ characterize the loss rate of the polariton and hot excitons, respectively, 
and $P$ represents of the strength of the laser pump. {Note that the parameters of $g_C$ , $g_R$ , and~$R$ in Equations~(\ref{eq.1}) and (\ref{eq.2}) have been
rescaled into the one-dimensional case by the width
$d$ of the nanowire thickness as ($g_C \rightarrow g_C /\sqrt{2\pi d}$,
$g_R \rightarrow g_R /\sqrt{2\pi d}$, and $R \rightarrow R /\sqrt{2\pi d}$). We remark that the model system described by Equations~(\ref{eq.1}) and (\ref{eq.2}) corresponds to an exciton--polariton Bose--Einstein condensate under nonresonant pumping, realized in a wire-shaped microcavity analogous to that implemented in Ref.~\cite{Wertz2010}. This geometry confines polaritons to a quasi-one-dimensional (1D) channel.}

In relevant experiments, the~values of the parameters can be set as follows:
$m=5\times10^{-5}m_{e}$, $g_{C}=0.475\mu \text{eV} \cdot \mu \text{m}$, $g_{R}=0.95 \mu \text{eV} \cdot \mu \text{m}$,
$R=2.24\times10^{-4} \mu \text{mps}^{-1}$,$\gamma_{R}=1.5\times\gamma_{C}=0.25\text{ps}^{-1}$, and
$P=1.1\text{ps}^{-1} \mu \text{m}^{-2}$. 
We study the quench dynamics in a repulsive one-dimensional exciton--polariton condensate, 
where both the nonlinearity parameter and the laser pump parameter undergo a sudden~quench.

For subsequent analysis, all variables need to be dimensionless in Equations (\ref{eq.1}) and (\ref{eq.2}).
In~the dimensionless process, we set $x^{\prime}=x/r_{h}$ and $t^{\prime}=t/\tau_{0}$, where $\tau_{0}=r_{h}/c_{s}=mr_{h}^{2}/\hbar$, $c_{s}=\hbar/m/r_{h}$ is the local sound velocity , and $r_{h}$ is the healing length in the condensate. As~such, Equations (\ref{eq.1}) and (\ref{eq.2}) can be rewritten into

\begin{eqnarray}
i\frac{\partial\psi}{\partial t} & =& \left[ -\frac{1}{2}\frac{\partial^{2}}{\partial x^{2}}+g\left|\psi\right|^{2}\right]\psi \nonumber\\
&+&  \left[\bar{g}_{R}\bar{n}_{R}+\frac{i}{2}\left[\bar{R}\bar{n}_{R}-\bar{\gamma}_{C}\right]\right]\psi, \label{eq.3} \\ 
\frac{\partial\bar{n}_{R}}{\partial t} & =& \bar{P}-\left(\bar{\gamma}_{R}+\bar{R}\left|\psi\right|^{2}\right)\bar{n}_{R}, \label{eq.4} 
\end{eqnarray}
with $mr_{h}^{2}g_{C}/\hbar^{2}\equiv g$, $mr_{h}^{2}g_{R}/\hbar^{2}\equiv \bar{g}_{R}$, $mr_{h}^{2}R/\hbar\equiv \bar{R}$, $mr_{h}^{2}\gamma_{C}/\hbar\equiv \bar{\gamma}_{C}$, $mr_{h}^{2}\gamma_{R}/\hbar\equiv \bar{\gamma}_{R}$, $n_{R}\equiv \bar{n}_{R}$, and $t^{\prime}\rightarrow t$ and $x^{\prime}\rightarrow x$ set in the above equations.  Furthermore we consider $\psi\left(x, t\right)=\psi^{\prime}\left(x, t\right) \exp \left[-i g n_{0}t\right]$. 
The resulting dimensionless equations take the form
\begin{eqnarray}
i\frac{\partial}{\partial t}\psi^{\prime}& =& \left[-\frac{1}{2}\frac{\partial^{2}}{\partial x^{2}}+g\left(\left|\psi^{\prime}\right|^{2}-n_{0}\right)\right]\psi^{\prime} \nonumber\\
&+&\left[\bar{g}_{R}\bar{n}_{R}+\frac{i}{2}\left(\bar{R}\bar{n}_{R}-\bar{\gamma}_{C}\right)\right]\psi^{\prime},  \label{eq.5} \\
\frac{\partial}{\partial t}\bar{n}_{R}& =& \bar{P}-\left(\bar{\gamma}_{R}+\bar{R}\left|\psi^{\prime}\right|^{2}\right)\bar{n}_{R}. \label{eq.6} 
\end{eqnarray}

Then, we restrict our analysis to the fast reservoir limit, characterized by the condition $\bar{\gamma}_{C} \ll \bar{\gamma}_{R}$.  
Expanding the reservoir dynamics to leading order, the~hot exciton reservoir density $n_R$ satisfies
\begin{equation}
\bar{n}_{R}=\frac{\bar{P}}{\bar{\gamma}_{R}}-\frac{\bar{P}\bar{R}}{\bar{\gamma}_{R}^{2}}\left|\psi^{\prime}\right|^{2}.  \label{eq.7} 
\end{equation}
Under the fast reservoir limit, the~dynamics of the condensate results from Equations~(\ref{eq.5}) and (\ref{eq.6}) reads
\begin{align}
i\frac{\partial}{\partial t}\psi^{\prime}+\frac{1}{2}\frac{\partial^{2}}{\partial x^{2}}\psi^{\prime}-g\left(\left|\psi^{\prime}\right|^{2}-n_{0}\right)\psi^{\prime}=P\left(\psi^{\prime}\right), \label{eq.8} 
\end{align}
where $P\left(\psi^{\prime}\right)$ represents the nonequilibrium nature, i.e.,~the intrinsic kinds of gain and loss of the polariton condensate, which is given by
\begin{eqnarray}
P\left(\psi^{\prime}\right) & =&\left[\bar{g}_{R}\bar{n}_{R}+\frac{i}{2}\left(\bar{R}\bar{n}_{R}-\bar{\gamma}_{C}\right)\right]\psi^{\prime}, \nonumber \\
 & =&\bar{g}_{R}\left(\frac{\bar{P}}{\bar{\gamma}_{R}}-\frac{\bar{P}\bar{R}}{\bar{\gamma}_{R}^{2}}\left|\psi^{\prime}\right|^{2}\right)\psi^{\prime}  \nonumber \\
 &+&\frac{i}{2}\left[\bar{R}\left(\frac{\bar{P}}{\bar{\gamma}_{R}}-\frac{\bar{P}\bar{R}}{\bar{\gamma}_{R}^{2}}\left|\psi^{\prime}\right|^{2}\right)-\bar{\gamma}_{C}\right]\psi^{\prime}. \label{eq.9} 
\end{eqnarray}
In what follows, we plan to investigate the quench dynamics and stability of dark solitons in repulsive one-dimensional exciton--polariton condensates undergoing a sudden quench simultaneously in the nonlinearity and pump parameters based on both analytically and numerically solving Equations~(\ref{eq.8}) and (\ref{eq.9}). {The emphasis and value of this work is to investigate how the relationship between the dissipation and pumping can affect the quench dynamics of the nonlinear excitations in a nonequilibrium condensate.}

\section{The Hamiltonian Approach and Dynamics of the Dark Soliton Induced by the Quenched~Interaction}\label{Sec3}

The goal of Section~\ref{Sec3} is to derive the equation of motion for the center of mass of the dark soliton by solving Equation~(\ref{eq.8}) subjected to time-dependent perturbations provided by Equation~(\ref{eq.9}) in the limit of a fast~reservoir. 

As a first step, we  initially consider the simple case where $P\left(\psi^\prime\right)=0$ in Equation~(\ref{eq.8}), i.e.,~in the absence of the open-dissipative. 
Meanwhile, we limit the case of $g>0$ corresponding to the repulsive interaction. There exists an exact dark soliton solution to Equation~(\ref{eq.8}), taking the form of
\begin{equation}
\psi^{\prime}=\sqrt{n_{0}}\left[i\frac{u}{c_{s}}+\sqrt{1-\frac{u^{2}}{c_{s}^{2}}}\tanh\left[\sqrt{1-\frac{u^{2}}{c_{s}^{2}}}\frac{x-ut}{\sqrt{2}r_{h}}\right]\right], \label{eq.10}
\end{equation}
with $n_{0}$ being the background density and $r_{h}$ being the healing length or coherence length. 
We recall that $c_{s}$ accounts for the speed of sound while $u$ denotes the velocity of the dark soliton itself. 
Now, we recall the fundamental properties of dark solitons. 
A soliton is often referred to as a “black” soliton when the velocity is zero. However, for~a moving (“gray”) soliton,
the minimum value of density $n_{\text{min}}$ increases in proportion to the square of the soliton's velocity, i.e.,~$n_{\text{min}}=u^{2}$. 

Next, we consider the open-dissipative properties as perturbations to the condensate and study the finite-amplitude  collective excitations in a homogeneous condensate as captured by $P\left(\psi^\prime\right)\neq 0$.
In other words, our aim is to investigate the quench dynamics of dark solitons in condensates with nonequilibrium properties. 
To the end, we plan to use the Hamiltonian approach~\cite{Yiling_Dynamics_2022,xu_darkbright_2019,xue_creation_2014,Cao2024,Cao2025} to solve  Equation~(\ref{eq.8}).  At~the heart of the Hamiltonian approach of quantum dynamics for dark solitons is the assumption that 
the parameters of the dark solitons supported by Equation~(\ref{eq.8})  become slow functions of time in the presence of perturbation in Equation~(\ref{eq.9}), i. e. $u\rightarrow u(t)$, while the functional form of the dark soliton in Equation~(\ref{eq.10}) remains unchanged. As~such, the~time-dependent
variation in the soliton's parameters can be analytically derived by the time evolution of the dark soliton's energy as follows,
\begin{equation}
\frac{dE}{dt}=\int dx\left(P^{*}\left(\psi^{\prime}\right)\frac{d}{dt}\psi^{\prime}+P\left(\psi^{\prime}\right)\frac{d}{dt}\psi^{\prime*}\right). \label{eq.11}
\end{equation}
In Equation~(\ref{eq.11}),  the excitation energy of the dark soliton can be calculated by
\begin{align}
E & =\frac{1}{2}\int dx\left(\left|\frac{\partial\psi^{\prime}}{\partial x}\right|^{2}+g\left(n_{0}-\left|\psi^{\prime}\right|^{2}\right)^{2}\right), \nonumber \\
 & =\frac{\sqrt{2}n_{0}\left(c_{s}^{2}-u^{2}\right)^{3/2}\left(1+2gr_{h}^{2}n_{0}\right)}{3c_{s}^{3}r_{h}}. \label{eq.12}
\end{align}
Before  proceeding, we establish methodological distinctions between our Hamiltonian approach and related analytical frameworks in Refs.~\cite{wouters_excitations_2007,carretero-gonzalez_kortewegvries_2017}:  (i) While Ref.~\cite{wouters_excitations_2007} investigates linear excitations and their spectral properties through Bogoliubov's theory within the linear response formalism, our work focuses on quench dynamics of nonlinear excitations—specifically dark solitons—exhibiting non-perturbative temporal evolution. (ii)  Both our study and Ref.~\cite{carretero-gonzalez_kortewegvries_2017} address dark soliton dynamics but~employ distinct approaches: Carretero-González~et~al. in Ref.~\cite{carretero-gonzalez_kortewegvries_2017} utilize reductive perturbation theory to derive a Korteweg--de Vries (KdV) equation with linear damping, valid under weak reservoir coupling where reservoir excitations remain infinitesimal.
 Conversely, we operate in the fast reservoir limit by directly solving the microscopic Equation~(\ref{eq.8}), capturing nonlinear soliton--reservoir interactions without perturbative approximations.

At the end of the analytical calculation, by~inserting Equations~(\ref{eq.9}) and (\ref{eq.12}) into Equation~(\ref{eq.11}), 
one can smoothly derive the equation of motion for the velocity of the dark soliton, which takes the form
\begin{align}
\frac{du}{dt} & =\frac{r_{h}}{3\sqrt{2}\left(1+2gr_{h}^{2}n_{0}\right)c_{s}\bar{\gamma}_{R}^{2}} \nonumber \\
 & \times\left\{ 6c_{s}^{2}u\bar{\gamma}_{R}\left(\bar{P}\bar{R}-\bar{\gamma}_{C}\bar{\gamma}_{R}\right)-2un_{0}\bar{P}\bar{R}^{2}\left(2u^{2}+c_{s}^{2}\right)\right\} .  \label{eq.13}
\end{align}

We find that, for~a certain~situation where $\bar{\gamma}_{C}\ll\bar{\gamma}_{R}$,
the trajectory of the dark soliton over long-term evolution indeed conforms to the form of Equation~(\ref{eq.13}). 
According to this formula, we understand that the quench dynamics of the dark solitons is solely dependent on 
the quench interaction strength and the open-dissipative nature of the reservoir system. 


Now, we are ready to investigate  the quench dynamics and stability of dark solitons in repulsive one-dimensional exciton--polariton condensates undergoing a sudden quench simultaneously in the nonlinearity and pump parameters based on both analytically and numerically solving Equations~(\ref{eq.8}) and (\ref{eq.9}) by designing two scenarios. The~first scenario involves exciton--polariton condensates undergoing a sudden quench of the nonlinearity parameter;
i.e., the~strength of the interaction changes from $g_{1}$ to $g_{2}$.
On the other hand, the~second scenario involves exciton--polariton condensates undergoing a sudden quench of the laser pump parameter; 
i.e., the~strength of the pump changes from $\bar{P}_{1}$ to $\bar{P}_{2}$. 
This can be achieved by either varying the coupling constant $g$ of the system or changing the laser pump $P$ in the experiment. 
The above expression or above result can also be illustrated in diagrammatic form, as~shown in Figures 1 and 2.

\begin{figure}[htb]
\centering
\includegraphics[width=0.85\columnwidth]{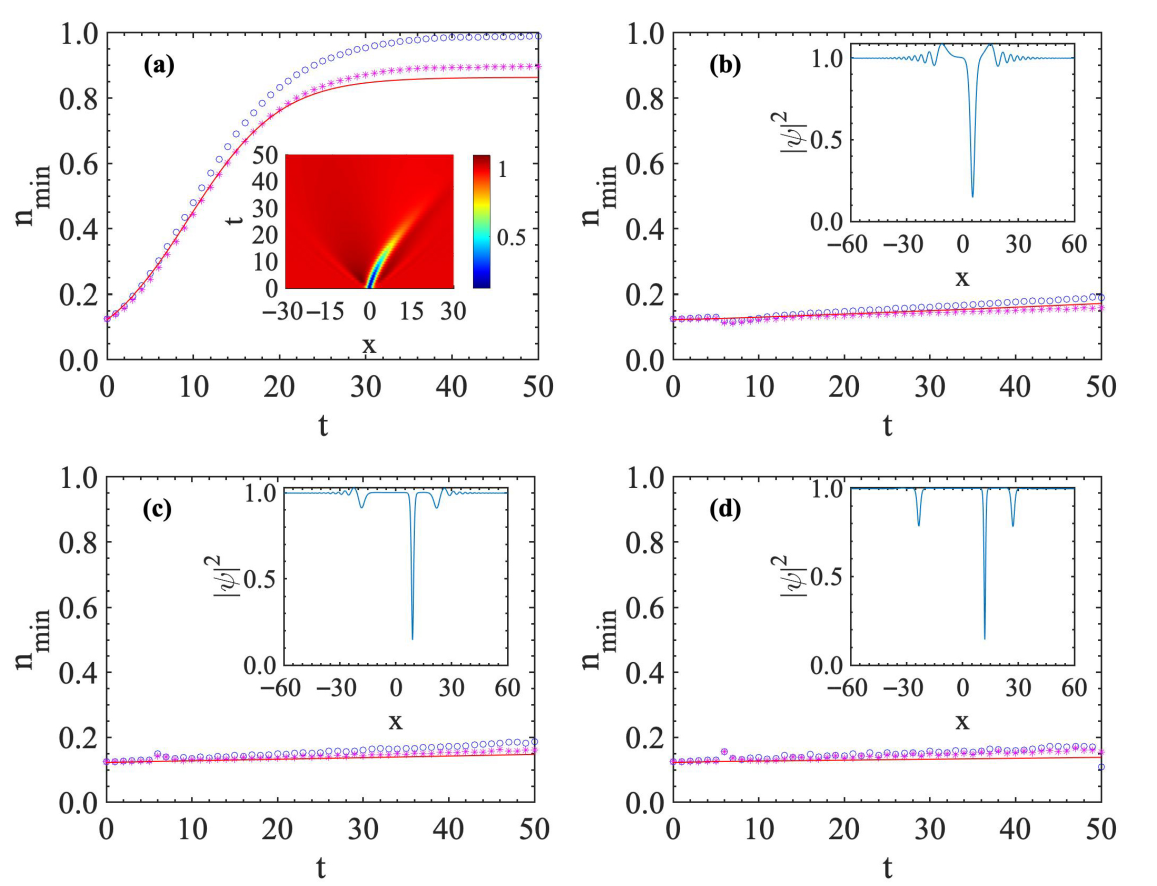}\\
\caption{Quench dynamics of the dark soliton with the initial velocity $u=0.35$ after the quench \linebreak  $g_{1}$ $\rightarrow$ $g_{2}$. 
The varied strength of the interaction after the quench $g_{2}=1,~0.5,~2,~4$ corresponds to (\textbf{a}), (\textbf{b}), (\textbf{c}), and~(\textbf{d}), respectively.
The other parameters are set as $g_{1}=1$, $\bar{g}_{R}=2/30$, $\bar{\gamma}_{C}=3/30$, $\bar{\gamma}_{R}=15/30$, $\bar{R}=1.5/30$,
and $\bar{P}=33/30$. 
The inset of (\textbf{a}) shows the evolution of the condensate density distribution $n=\left|\psi^{\prime}\right|^{2}$.
The inset of (\textbf{b}--\textbf{d}) shows the profile of the dark soliton with $t=20$. 
The analytical solution for the minimum value of density $n_{\text{min}}$ using Equation~(\ref{eq.13}) is plotted as a red solid line. 
The numerical solution obtained from evolution Equation~(\ref{eq.8}) is plotted with a red star line. 
The numerical solution obtained from real-time evolution Equations~(\ref{eq.5}) and (\ref{eq.6})  is plotted with blue~circle.}
\label{F1}
\end{figure}

We revisit the scenario of non-quenching dynamics under nonresonant pumping, where the interaction strength remains consistent before and after the non-quenching process—consistent with the setup described in Ref.~\cite{smirnov_dynamics_2014} (as illustrated in Figure~\ref{F1}a). As~depicted, in~an open-dissipative system, dark solitons eventually decay after persisting for a finite duration, with~their lifetime modulated by dissipation parameters. The~dynamical trajectory and lifetime of such dark solitons can be obtained via real-time evolution of the Gross--Pitaevskii~equation.

Figure~\ref{F1}b--d present the time evolution of the minimum condensate density associated with the dark soliton, corresponding to $g_{2}=\frac{1}{2}g_{1}$ , $g_{2}=2g_{1}$, and $g_{2}=4g_{1}$, respectively. Key observations from Figure~\ref{F1} are summarized as follows: When the post-quench interaction strength is halved relative to the initial value (Figure \ref{F1}b), shock waves emerge during dark soliton evolution. For~a post-quench strength doubled relative to the initial value (Figure \ref{F1}c), the~soliton splits into two distinct solitons, accompanied by the formation of a shock wave. In~contrast, when the post-quench strength is quadrupled (Figure \ref{F1}d), the~soliton splits during evolution, producing a pair of asymmetric solitons without any shock wave formation. Soliton splitting occurs due to changes in the interaction strength, which alter the dynamical phase of the soliton. This causes the initial soliton to split into two faster-moving solitons, while the original soliton becomes narrower.

\begin{figure}[htb]
\centering
\includegraphics[width=1\columnwidth]{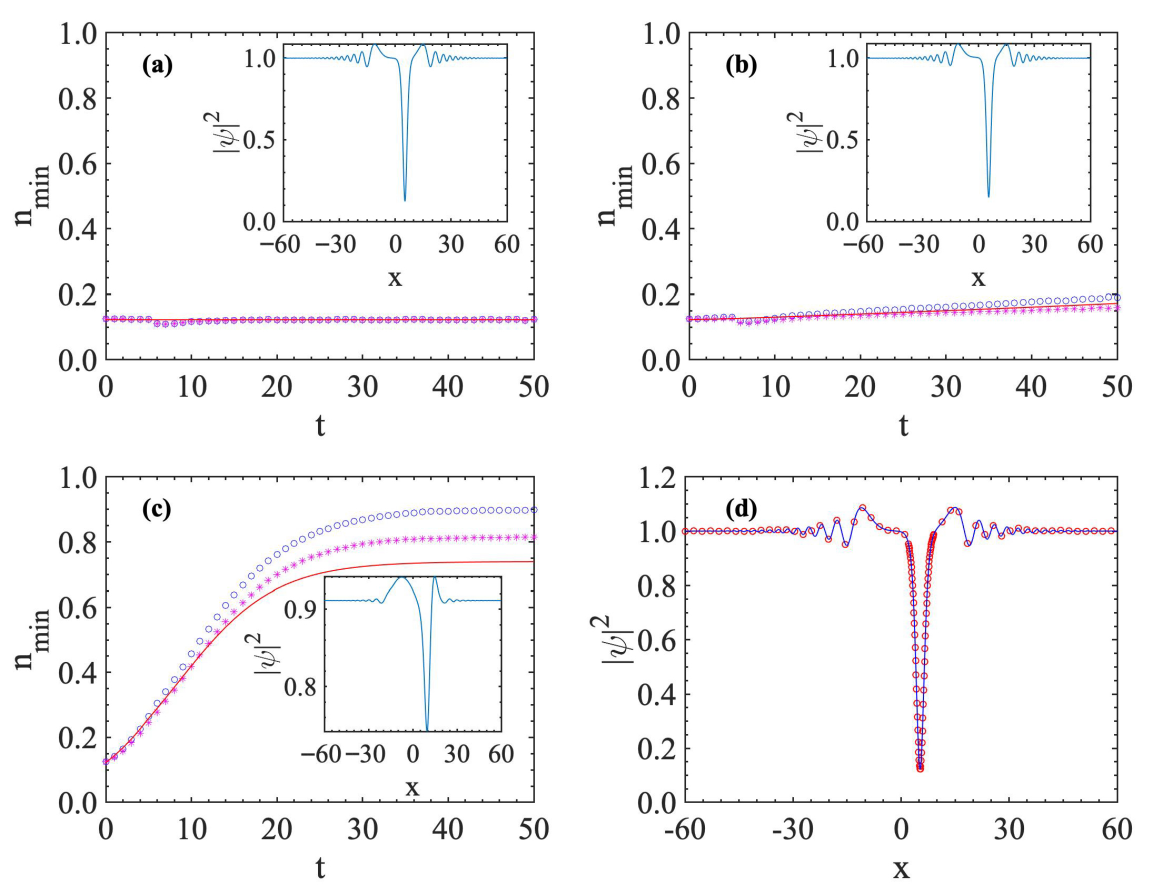}\\
\caption{Quench dynamics of a 1D dark soliton with the initial velocity $u=0.35$ under quenching laser pumping intensity. 
In (\textbf{a},\textbf{d}) we show the quench dynamics of the dark soliton in the absence of the open-dissipative with $g_{1}=1$ and $g_{2}=0.5$.
In (\textbf{b},\textbf{c}) we show the quench dynamics of the dark soliton in the presence of the open-dissipative.
The varied strength of the pump after the quench $\bar{P}_{2}=27,~30$ corresponds to (\textbf{b}) and (\textbf{c}), respectively. 
The parameters are chosen as $g=1$, $\bar{g}_{R}=2$, $\bar{\gamma}_{C}=3$, $\bar{\gamma}_{R}=15$, $\bar{R}=1.5$,
and $\bar{P}_{1}=33$.
In (\textbf{a}--\textbf{c}), the meaning represented by different lines is the same as in Figure 1. 
In (\textbf{d}) the numerical solution  is plotted with a red~circle. The analytical solution is plotted as a blue solid line.}
\label{F2}
\end{figure}

Figure~\ref{F2}a illustrates the quench dynamics of a 1D dark soliton in a conventional Bose--Einstein condensate (BEC) via quenching the interaction strength, i.e.,~in the absence of open-dissipative effects. In~this conservative system, the~velocity of the dark soliton remains constant during propagation. This figure validates the correctness of our scheme by demonstrating the reproducibility of dark soliton quench dynamics in conservative regimes. 
By comparing with Figure~\ref{F1}b, we observe that while a shock wave still emerges for $g_{2}=\frac{1}{2}g_{1}$, the~presence of dissipation reduces the lifetime of both the soliton and the shock wave. Notably, Figure~\ref{F2}d shows that numerical solutions of Equations~(\ref{eq.5}) and (\ref{eq.6}) (plotted as red circles) agree well with those obtained from solving Equation~(\ref{eq.8}) (plotted as a blue curve) within the allowable error range.
Figure~\ref{F2}b,c present the quench dynamics of a 1D dark soliton under laser pumping intensity quenching, corresponding to  
$\bar{P}_{2}=27$ and $\bar{P}_{2}=30$, respectively. The~insets in (a), (b), and~(c) display the density profiles at $t=20$. These results reveal that shock waves persist during dark soliton propagation even under laser pumping intensity~quenching.


\section{Discussion and~Conclusions}\label{Sec4}
In summary, we have investigated the quench dynamics of dark solitons in a polariton condensate under nonresonant pumping. The~dynamics are studied through independent quenches of the interaction strength and the pump laser intensity. Using a Hamiltonian approach, we have derived analytical expressions governing the dark soliton's equation of motion for its velocity. During~the evolution, phenomena such as splitting of the initial soliton, generation of {asymmetric} solitons, and~formation of shock waves may occur under specific scenarios. Our numerical solutions of the modified open-dissipative Gross--Pitaevskii equation confirm these analytical~findings.

\begin{acknowledgments}
We thank Ying Hu, Yapeng Zhang, Xuzhen Cao, Shujie Cheng, and~Biao Wu for stimulating discussions and useful~help.
This work was supported  by the National Natural Science Foundation of China (Grants No. 12574301) and the Zhejiang Provincial Natural  Science Foundation (Grant No. LZ25A040004).
The article has been published in Symmetry (ISSN 2073-8994).
Citation: Jia, C.; Liang, Z. Quench Dynamics and Stability of Dark Solitons in Exciton–Polariton Condensates. Symmetry 2025, 17, 1482.
https://doi.org/10.3390/sym17091482
\end{acknowledgments}

\bibliographystyle{unsrt}
\normalem
\bibliography{quench}

\begin{thebibliography}{10}

\bibitem{Baumberg_Spontaneous_2008}
J.~J. Baumberg, A.~V. Kavokin, S.~Christopoulos, A.~J.~D. Grundy, R.~Butt\'e,
  G.~Christmann, D.~D. Solnyshkov, G.~Malpuech, G.~Baldassarri H\"oger~von
  H\"ogersthal, E.~Feltin, J.-F. Carlin, and N.~Grandjean.
\newblock Spontaneous polarization buildup in a room-temperature polariton
  laser.
\newblock {\em Phys. Rev. Lett.}, 101:136409, Sep 2008.

\bibitem{Yiling_Dynamics_2022}
Yiling Zhang, Chunyu Jia, and Zhaoxin Liang.
\newblock Dynamics of two dark solitons in a polariton condensate.
\newblock {\em Chin. Phys. Lett.}, 39(2):020501--020501, 2022.

\bibitem{kasprzak_boseeinstein_2006}
J.~Kasprzak, M.~Richard, S.~Kundermann, A.~Baas, P.~Jeambrun, J.~M.~J. Keeling,
  F.~M. Marchetti, M.~H. Szymańska, R.~André, J.~L. Staehli, V.~Savona, P.~B.
  Littlewood, B.~Deveaud, and Le~Si Dang.
\newblock Bose–{Einstein} condensation of exciton polaritons.
\newblock {\em Nature}, 443(7110):409--414, 2006.

\bibitem{balili_bose-einstein_2007}
R.~Balili, V.~Hartwell, D.~Snoke, L.~Pfeiffer, and K.~West.
\newblock Bose-{Einstein} {Condensation} of {Microcavity} {Polaritons} in a
  {Trap}.
\newblock {\em Science}, 316(5827):1007--1010, 2007.

\bibitem{deng_spatial_2007}
Hui Deng, Glenn~S. Solomon, Rudolf Hey, Klaus~H. Ploog, and Yoshihisa Yamamoto.
\newblock Spatial {Coherence} of a {Polariton} {Condensate}.
\newblock {\em Physical Review Letters}, 99(12):126403, 2007.

\bibitem{amo_collective_2009}
A.~Amo, D.~Sanvitto, F.~P. Laussy, D.~Ballarini, E.~del Valle, M.~D. Martin,
  A.~Lemaître, J.~Bloch, D.~N. Krizhanovskii, M.~S. Skolnick, C.~Tejedor, and
  L.~Viña.
\newblock Collective fluid dynamics of a polariton condensate in a
  semiconductor microcavity.
\newblock {\em Nature}, 457(7227):291--295, 2009.

\bibitem{schneider_electrically_2013}
Christian Schneider, Arash Rahimi-Iman, Na~Young Kim, Julian Fischer, Ivan~G.
  Savenko, Matthias Amthor, Matthias Lermer, Adriana Wolf, Lukas Worschech,
  Vladimir~D. Kulakovskii, Ivan~A. Shelykh, Martin Kamp, Stephan Reitzenstein,
  Alfred Forchel, Yoshihisa Yamamoto, and Sven Höfling.
\newblock An electrically pumped polariton laser.
\newblock {\em Nature}, 497(7449):348--352, 2013.

\bibitem{christopoulos_room-temperature_2007}
S.~Christopoulos, G.~Baldassarri~Höger von Högersthal, A.~J.~D. Grundy, P.~G.
  Lagoudakis, A.~V. Kavokin, J.~J. Baumberg, G.~Christmann, R.~Butté,
  E.~Feltin, J.-F. Carlin, and N.~Grandjean.
\newblock Room-{Temperature} {Polariton} {Lasing} in {Semiconductor}
  {Microcavities}.
\newblock {\em Physical Review Letters}, 98(12):126405, 2007.

\bibitem{ruprecht_time-dependent_1995}
P.~A. Ruprecht, M.~J. Holland, K.~Burnett, and Mark Edwards.
\newblock Time-dependent solution of the nonlinear {Schrödinger} equation for
  {Bose}-condensed trapped neutral atoms.
\newblock {\em Physical Review A}, 51(6):4704--4711, 1995.

\bibitem{burger_dark_1999}
S.~Burger, K.~Bongs, S.~Dettmer, W.~Ertmer, K.~Sengstock, A.~Sanpera, G.~V.
  Shlyapnikov, and M.~Lewenstein.
\newblock Dark {Solitons} in {Bose}-{Einstein} {Condensates}.
\newblock {\em Physical Review Letters}, 83(25):5198--5201, 1999.

\bibitem{WLZhang_Coupled_2012}
W.L. Zhang and Y.J. Rao.
\newblock Coupled polariton solitons in semiconductor microcavities with a
  double-well potential.
\newblock {\em Chaos, Solitons \& Fractals}, 45(4):373--377, 2012.

\bibitem{xue_creation_2014}
Yan Xue and Michał Matuszewski.
\newblock Creation and {Abrupt} {Decay} of a {Quasistationary} {Dark} {Soliton}
  in a {Polariton} {Condensate}.
\newblock {\em Physical Review Letters}, 112(21):216401, 2014.

\bibitem{Sich_Soliton_2016}
Maksym Sich, Dmitry~V. Skryabin, and Dmitry~N. Krizhanovskii.
\newblock Soliton physics with semiconductor exciton–polaritons in confined
  systems.
\newblock {\em Comptes Rendus Physique}, 17(8):908--919, 2016.
\newblock Polariton physics / Physique des polaritons.

\bibitem{Kartashov_latticesolitons_2016}
Yaroslav~V. Kartashov and Dmitry~V. Skryabin.
\newblock Two-dimensional lattice solitons in polariton condensates with
  spin-orbit coupling.
\newblock {\em Opt. Lett.}, 41(21):5043--5046, Nov 2016.

\bibitem{Walker_Dark_2017}
P.~M. Walker, L.~Tinkler, B.~Royall, D.~V. Skryabin, I.~Farrer, D.~A. Ritchie,
  M.~S. Skolnick, and D.~N. Krizhanovskii.
\newblock Dark solitons in high velocity waveguide polariton fluids.
\newblock {\em Phys. Rev. Lett.}, 119:097403, Aug 2017.

\bibitem{xu_darkbright_2019}
Xingran Xu, Lei Chen, Zhidong Zhang, and Zhaoxin Liang.
\newblock Dark–bright solitons in spinor polariton condensates under
  nonresonant pumping.
\newblock {\em Journal of Physics B: Atomic, Molecular and Optical Physics},
  52(2):025303, 2019.

\bibitem{Septembre_Soliton_2024}
I.~Septembre, I.~Foudjo, V.~Develay, T.~Guillet, S.~Bouchoule, J.~Z\'u\~niga
  P\'erez, D.~D. Solnyshkov, and G.~Malpuech.
\newblock Soliton formation in an exciton-polariton condensate at a bound state
  in the continuum.
\newblock {\em Phys. Rev. B}, 109:205302, May 2024.

\bibitem{HuJunwei_Darksolitoncloning_2024}
Junwei Hu, Kun Zhang, Muhammad Idrees, Hui-jun Li, Ji~Lin, and Alexey Kavokin.
\newblock Dark soliton cloning in exciton-polariton condensates.
\newblock {\em Phys. Rev. B}, 110:155112, Oct 2024.

\bibitem{krizhanovskii_effect_2010}
D.~N. Krizhanovskii, D.~M. Whittaker, R.~A. Bradley, K.~Guda, D.~Sarkar,
  D.~Sanvitto, L.~Vina, E.~Cerda, P.~Santos, K.~Biermann, R.~Hey, and M.~S.
  Skolnick.
\newblock Effect of {Interactions} on {Vortices} in a {Nonequilibrium}
  {Polariton} {Condensate}.
\newblock {\em Physical Review Letters}, 104(12):126402, 2010.

\bibitem{lagoudakis_quantized_2008}
K.~G. Lagoudakis, M.~Wouters, M.~Richard, A.~Baas, I.~Carusotto, R.~André,
  Le~Si Dang, and B.~Deveaud-Plédran.
\newblock Quantized vortices in an exciton–polariton condensate.
\newblock {\em Nature Physics}, 4(9):706--710, 2008.

\bibitem{ma_creation_2017}
Xuekai Ma, Oleg~A. Egorov, and Stefan Schumacher.
\newblock Creation and {Manipulation} of {Stable} {Dark} {Solitons} and
  {Vortices} in {Microcavity} {Polariton} {Condensates}.
\newblock {\em Physical Review Letters}, 118(15):157401, 2017.

\bibitem{rubo_half_2007}
Yuri~G. Rubo.
\newblock Half {Vortices} in {Exciton} {Polariton} {Condensates}.
\newblock {\em Physical Review Letters}, 99(10):106401, 2007.

\bibitem{cuevas_nonlinear_2011}
J.~Cuevas, A.~S. Rodrigues, R.~Carretero-González, P.~G. Kevrekidis, and D.~J.
  Frantzeskakis.
\newblock Nonlinear excitations, stability inversions, and dissipative dynamics
  in quasi-one-dimensional polariton condensates.
\newblock {\em Physical Review B}, 83(24):245140, 2011.

\bibitem{pinsker_approximate_2015}
Florian Pinsker.
\newblock Approximate solutions for half-dark solitons in spinor
  non-equilibrium {Polariton} condensates.
\newblock {\em Annals of Physics}, 362:726--738, 2015.

\bibitem{wang_dark_2003}
Shun-Jin Wang, Cheng-Long Jia, Dun Zhao, Hong-Gang Luo, and Jun-Hong An.
\newblock Dark and bright solitons in a quasi-one-dimensional {Bose}-{Einstein}
  condensate.
\newblock {\em Physical Review A}, 68(1):015601, 2003.

\bibitem{timmermans_feshbach_1999}
E~Timmermans.
\newblock Feshbach resonances in atomic {Bose}–{Einstein} condensates.
\newblock {\em Physics Reports}, 315(1-3):199--230, 1999.

\bibitem{inouye_observation_1998}
S.~Inouye, M.~R. Andrews, J.~Stenger, H.-J. Miesner, D.~M. Stamper-Kurn, and
  W.~Ketterle.
\newblock Observation of {Feshbach} resonances in a {Bose}–{Einstein}
  condensate.
\newblock {\em Nature}, 392(6672):151--154, 1998.

\bibitem{gamayun_fate_2015}
O.~Gamayun, Yu.~V. Bezvershenko, and V.~Cheianov.
\newblock Fate of a gray soliton in a quenched {Bose}-{Einstein} condensate.
\newblock {\em Physical Review A}, 91(3):031605, 2015.

\bibitem{chen_observation_2020}
Cheng-An Chen and Chen-Lung Hung.
\newblock Observation of {Universal} {Quench} {Dynamics} and {Townes} {Soliton}
  {Formation} from {Modulational} {Instability} in {Two}-{Dimensional} {Bose}
  {Gases}.
\newblock {\em Physical Review Letters}, 125(25):250401, 2020.

\bibitem{drescher_real-space_2019}
Moritz Drescher, Manfred Salmhofer, and Tilman Enss.
\newblock Real-space dynamics of attractive and repulsive polarons in
  {Bose}-{Einstein} condensates.
\newblock {\em Physical Review A}, 99(2):023601, 2019.

\bibitem{deng_tuning_2017}
Tian-Shu Deng, Wei Zhang, and Wei Yi.
\newblock Tuning {Feshbach} resonances in cold atomic gases with interchannel
  coupling.
\newblock {\em Physical Review A}, 96(5):050701, 2017.

\bibitem{Hamid_Solitonsplitting_1997}
Hamid Hatami-Hanza, P.L. Chu, Boris~A. Malomed, and G.D. Peng.
\newblock Soliton compression and splitting in double-core nonlinear optical
  fibers.
\newblock {\em Optics Communications}, 134(1):59--65, 1997.

\bibitem{porsezian_soliton_2015}
K.~PORSEZIAN and R.~VASANTHA~JAYAKANTHA RAJA.
\newblock Soliton fission and supercontinuum generation in photonic crystal
  fibre for optical coherence tomography application.
\newblock {\em Pramana}, 85(5):993--1007, nov 2015.

\bibitem{wales_splitting_2020}
Oliver~J. Wales, Ana Rakonjac, Thomas~P. Billam, John~L. Helm, Simon~A.
  Gardiner, and Simon~L. Cornish.
\newblock Splitting and recombination of bright-solitary-matter waves.
\newblock {\em Communications Physics}, 3(1):51, mar 2020.

\bibitem{Li_Merging_2021}
Xin Li, Peng Gao, Zhan-Ying Yang, and Wen-Li Yang.
\newblock Merging and splitting dynamics between two bright solitons in dipolar
  bose–einstein condensates*.
\newblock {\em Chinese Physics B}, 30(12):120501, dec 2021.

\bibitem{Simmons_What_2020}
S.~A. Simmons, F.~A. Bayocboc, J.~C. Pillay, D.~Colas, I.~P. McCulloch, and
  K.~V. Kheruntsyan.
\newblock What is a quantum shock wave?
\newblock {\em Phys. Rev. Lett.}, 125:180401, Oct 2020.

\bibitem{jia_quenched_2021}
Rui-Yu Jia, Ping-Ping Fang, Chao Gao, and Ji~Lin.
\newblock Quenched solitons and shock waves in {Bose}-{Einstein} condensates.
\newblock {\em Acta Physica Sinica}, 70(18):180303--1, 2021.
\newblock Num Pages: 180303-8.

\bibitem{wang_generation_2024}
Jin-Ling Wang, Kun Zhang, Ji~Lin, and Hui-Jun Li.
\newblock Generation and modulation of shock waves in two-dimensional polariton
  condensates.
\newblock {\em Acta Physica Sinica}, 73(11):119601--1, 2024.
\newblock Num Pages: 119601-9.

\bibitem{Kulikov_Shock_2003}
Igor Kulikov and Michail Zak.
\newblock Shock waves in a bose-einstein condensate.
\newblock {\em Phys. Rev. A}, 67:063605, Jun 2003.

\bibitem{Damski_Formation_2004}
Bogdan Damski.
\newblock Formation of shock waves in a bose-einstein condensate.
\newblock {\em Phys. Rev. A}, 69:043610, Apr 2004.

\bibitem{Kamchatnov_Dissipationless_2004}
A.~M. Kamchatnov, A.~Gammal, and R.~A. Kraenkel.
\newblock Dissipationless shock waves in bose-einstein condensates with
  repulsive interaction between atoms.
\newblock {\em Phys. Rev. A}, 69:063605, Jun 2004.

\bibitem{Hoefer_Dispersive_2006}
M.~A. Hoefer, M.~J. Ablowitz, I.~Coddington, E.~A. Cornell, P.~Engels, and
  V.~Schweikhard.
\newblock Dispersive and classical shock waves in bose-einstein condensates and
  gas dynamics.
\newblock {\em Phys. Rev. A}, 74:023623, Aug 2006.

\bibitem{wan_dispersive_2007}
Wenjie Wan, Shu Jia, and Jason~W. Fleischer.
\newblock Dispersive superfluid-like shock waves in nonlinear optics.
\newblock {\em Nature Physics}, 3(1):46--51, jan 2007.

\bibitem{Ghofraniha_Shock_2012}
N.~Ghofraniha, S.~Gentilini, V.~Folli, E.~DelRe, and C.~Conti.
\newblock Shock waves in disordered media.
\newblock {\em Phys. Rev. Lett.}, 109:243902, Dec 2012.

\bibitem{nuno_vectorial_2019}
J.~Nuño, C.~Finot, G.~Xu, G.~Millot, M.~Erkintalo, and J.~Fatome.
\newblock Vectorial dispersive shock waves in optical fibers.
\newblock {\em Communications Physics}, 2(1):138, nov 2019.

\bibitem{Jang_Shock_2023}
Heesuk Jang, Hajun Song, Hae~Seog Koh, Taehyun Yoon, and Yong~Joon Kwon.
\newblock Shock wave generation in water by nanosecond pulse laser irradiation
  with 1064 and 2940 nm wavelengths.
\newblock {\em Optics \& Laser Technology}, 167:109670, 2023.

\bibitem{Yin_Investigation_2023}
Caiyu Yin, Haiting Yu, Zeyu Jin, Jingxi Liu, Wei Huang, and Shijie Wu.
\newblock Investigation of shock wave propagation and water cavitation in a
  water-filled double plate subjected to underwater blast.
\newblock {\em International Journal of Mechanical Sciences}, 253:108400, 2023.

\bibitem{Hang_Accessing_2023}
Chao Hang, Zhengyang Bai, Weibin Li, Anatoly~M. Kamchatnov, and Guoxiang Huang.
\newblock Accessing and manipulating dispersive shock waves in a nonlinear and
  nonlocal rydberg medium.
\newblock {\em Phys. Rev. A}, 107:033503, Mar 2023.

\bibitem{Qin_Shock-wave_2024}
Lu~Qin, Chao Hang, Guoxiang Huang, and Weibin Li.
\newblock Shock-wave generation and propagation in dissipative and nonlocal
  nonlinear rydberg media.
\newblock {\em Phys. Rev. A}, 110:013703, Jul 2024.

\bibitem{wouters_excitations_2007}
Michiel Wouters and Iacopo Carusotto.
\newblock Excitations in a {Nonequilibrium} {Bose}-{Einstein} {Condensate} of
  {Exciton} {Polaritons}.
\newblock {\em Physical Review Letters}, 99(14):140402, 2007.

\bibitem{carretero-gonzalez_kortewegvries_2017}
R.~Carretero-González, J.~Cuevas-Maraver, D.J. Frantzeskakis, T.P. Horikis,
  P.G. Kevrekidis, and A.S. Rodrigues.
\newblock A {Korteweg}–de {Vries} description of dark solitons in polariton
  superfluids.
\newblock {\em Physics Letters A}, 381(45):3805--3811, 2017.

\bibitem{Carusotto2013}
Iacopo Carusotto and Cristiano Ciuti.
\newblock Quantum fluids of light.
\newblock {\em Rev. Mod. Phys.}, 85:299--366, Feb 2013.

\bibitem{Wertz2010}
E.~Wertz, L.~Ferrier, D.~D. Solnyshkov, R.~Johne, D.~Sanvitto, A.~Lema{\^i}tre,
  I.~Sagnes, R.~Grousson, A.~V. Kavokin, P.~Senellart, G.~Malpuech, and
  J.~Bloch.
\newblock Spontaneous formation and optical manipulation of extended polariton
  condensates.
\newblock {\em Nat. Phys.}, 6(11):860--864, Nov 2010.

\bibitem{Cao2024}
Xuzhen Cao, Chunyu Jia, Ying Hu, and Zhaoxin Liang.
\newblock Nonlinear thouless pumping of solitons across an impurity.
\newblock {\em Phys. Rev. A}, 110:013305, Jul 2024.

\bibitem{Cao2025}
Xuzhen Cao, Chunyu Jia, Hao Lyu, Ying Hu, and Zhaoxin Liang.
\newblock Transport of vector solitons in spin-dependent nonlinear thouless
  pumps.
\newblock {\em Phys. Rev. A}, 111:023329, Feb 2025.

\bibitem{smirnov_dynamics_2014}
Lev~A. Smirnov, Daria~A. Smirnova, Elena~A. Ostrovskaya, and Yuri~S. Kivshar.
\newblock Dynamics and stability of dark solitons in exciton-polariton
  condensates.
\newblock {\em Physical Review B}, 89(23):235310, 2014.

\end{thebibliography}

\end{document}